\documentclass[preprint,showpacs,preprintnumbers,amsmath,amssymb]{revtex4}

\usepackage{graphicx}
\usepackage{dcolumn}
\usepackage{bm}

\begin{document}
\preprint{APS/123-QED}

\title{Itinerant electron metamagnetism in LaCo$_9$Si$_4$}

\author{H. Michor}\email{michor@ifp.tuwien.ac.at}
\author{M. El-Hagary}
\author{M. Della\,Mea}
\author{M.W. Pieper}
\author{M. Reissner}
\author{G. Hilscher}
\affiliation{Institut f\"ur Festk\"orperphysik, T.U. Wien, 
Wiedner Hauptstrasse 8--10, A-1040 Wien, Austria}

\author{S. Khmelevskyi}
\author{P. Mohn}
\author{G. Schneider}
\affiliation{Center for Computational Material Science, T.U. Wien, 
A-1040 Wien, Austria}

\author{G. Giester}
\affiliation{Institut f\"ur Mineralogie und Kristallographie, Universit\"at Wien,
A-1090 Wien, Austria}

\author{P. Rogl}
\affiliation{Institut f\"ur Physikalische Chemie, Universit\"at Wien,
W\"ahringerstrasse 42,  A-1090 Wien, Austria}

\date{Nov. 5, 2003}

\begin{abstract}
The strongly exchange enhanced Pauli paramagnet LaCo$_9$Si$_4$ is found to 
exhibit an itinerant metamagnetic phase transition with indications for 
metamagnetic quantum criticality. 
Our investigation comprises magnetic, specific heat, and NMR
measurements as well as ab-initio electronic structure calculations. 
The critical field is about 3.5\,T for $H||c$ and 6\,T for $H\bot c$, which is the lowest value ever found for rare earth intermetallic compounds. 
In the ferromagnetic state there appears a moment of about 0.2\,$\mu_B$/Co at the 
$16k$ Co-sites, but sigificantly smaller moments at the $4d$ and $16l$ Co-sites.
\end{abstract}

\pacs{61.66.-f, 65.40.-b, 71.20.-b, 75.20.-g}

\maketitle

Reports on superconductivity on the border of itinerant
electron ferromagnetism in UGe$_2$ and ZrZn$_2$\cite{saxena,pfleiderer}
attracted considerable interest on quantum critical phenomena 
in ferromagnetic materials and motivated the search for new materials
which are in the vicinity of a ferromagnetic (FM) quantum critical point 
at ambient pressure. 
A fascinating system which may fit in this scenario is the solid
solution LaCo$_{13-x}$Si$_x$ where ferromagnetism vanishes near
the stoichiometric composition LaCo$_9$Si$_4$ (see Refs.~\cite{rao1,rao2,huang}).
Notwithstanding the qualitative agreement between the initial reports there
have been rather inconsistent results for the Curie temperature of 
LaCo$_9$Si$_4$ with $T_C\simeq 900$\,K and $T_C\simeq 40$\,K in 
Refs.~\cite{rao1,huang}. 
Our reinvestigation of LaCo$_{13-x}$Si$_x$ in the vicinity of the 
stoichiometric 1-9-4 composition revealed a monotonous decrease of the 
Curie temperature with $T_C\simeq 79$\,K and 36\,K for $x=3.8$ and 3.9, 
respectively, and the absence of FM order in well annealed 
single phase LaCo$_9$Si$_4$~\cite{El-Hagary}. 

In this letter we show a full crystallographic characterization of the 
true ternary compound LaCo$_9$Si$_4$ and present convincing evidence for a magnetic 
instability  at relatively low fields of a few Tesla, namely the occurrence  
of itinerant electron metamagnetism (IEMM).

Polycrystalline samples LaCo$_9$Si$_4$ were synthesized by high frequency 
induction melting of metal ingots (La 4N, Co 4.5N and Si 6N) 
and subsequent heat treatment at 1050$^{\circ}$C for 10 days. The phase purity 
and composition has been verified by electron microprobe studies. 
\begin{table}[b]
\begin{tabular}{lcc}
   \hline\hline
 Atom & Wyckoff p. &  Coordinates \\
  \hline
\multicolumn{3}{c}{$a = 0.7833(1)$\,nm; $c = 1.15657(2)$\,nm}  \\
 \hline
   La & $4a$   & $(0,0,\frac{1}{4})$ \\
  Co(1) & $16k$   &  $(x,y,0)$  \\
 &  & $x=0.06957(6)$; $y=0.20078(5)$  \\
  Co(2) & $16l$   & $(x,x+\frac{1}{2},z)$ \\
 & & $x=0.62708(4)$; $z=0.17994(4)$ \\
  Co(3) & $4d$  & $(0,\frac{1}{2},0)$ \\
  Si & $16l$ & $(x,x+\frac{1}{2},z)$ \\
 &  & $x=0.17013(8)$; $z=0.12080(8)$ \\
\hline
\hline
\end{tabular}
\caption{X-Ray Single Crystal Data for LaCo$_9$Si$_4$; Space Group
$I4/mcm$; No.140}
\end{table}
The room temperature structure investigation has been performed on a small 
single crystal ($56\times 72\times 56\mu $m$^3$) on a four circle 
Nonius Kappa diffractometer equipped with a CCD area detector.  
430 reflections  $> 4\sigma (F_0) $ out of 480 have been used for the
structure refinement.  
All details of the applied methodology were recently summarized
in context with our structure investigation on Ce- and LaNi$_9$Si$_4$
\cite{michor}. 
Analogous to LaNi$_9$Si$_4$ single crystal X-ray diffraction reveals for 
stoichiometric LaCo$_9$Si$_4$ a fully  ordered tetragonal LaFe$_9$Si$_4$-type 
structure~\cite{tang} (NaZn$_{13}$-derivative with {\it space group} $I4/mcm$).  
The occupancies of all crystallographic sites have been refined but did not 
reveal any significant deviations from stoichiometry. 
Refining anisotropic thermal displacement factors in the final run yielded 
$R$-values as low as 0.02 confirming the structural model with full atom order.
The results of the structure determination are summarized in Table I.
Being a fully ordered ternary compound LaCo$_9$Si$_4$ takes an  
exceptional position within the solid solution LaCo$_{13-x}$Si$_x$
which is corroborated by the residual resistivity values showing a
clear minimum at $x=4$ with $\rho_0\simeq 17\mu\Omega$cm 
(see Ref.~\cite{El-Hagary}) and by the 
limited solid solubility just around the stoichiometric
composition where small off-stoichiometries e.g.\ $x=4.05$ or
$x=3.95$ yielded inhomogeneous samples with secondary phases.

The observation of FM order in LaCo$_{13-x}$Si$_x$  
approaching zero temperature approximately at the stoichiometric 
composition 1-9-4 suggests that LaCo$_9$Si$_4$ may be at or nearby
a FM quantum critical point. This is supported by a rather large 
Sommerfeld value $\gamma\simeq 200$\,mJ/mol\,K$^2$ of LaCo$_9$Si$_4$ 
which is the maximum value within the solid solution.
The temperature dependencies of the electrical resistivity and magnetic 
susceptibility measured in external magnetic fields $\mu_0H<3$\,T 
in fact indicate a spin fluctuation regime~\cite{El-Hagary}. 
The dc magnetic susceptibility $\chi (T)$ shown in Fig.~1 exhibits 
a pronounced maximum at about 20\,K. The low temperature susceptibility 
$\chi_0$ of a large randomly oriented polycrystalline sample measured
with a vibrating sample magnetometer (VSM) at 1\,T 
(filled symbols in Fig.~1) amounts to $\chi_0\simeq 0.051$\,cm$^3$/mol. 
SQUID measurements performed on small c-axis oriented textured specimens
($<1$\,mg) reveal a significant anisotropy of the paramagnetic susceptibility 
yielding $\chi_0(H||c)\sim 0.07$\,cm$^3$/mol and 
$\chi_0(H\bot c)\simeq 0.04$\,cm$^3$/mol (open symbols in Fig.~1).

\begin{figure}[]
\includegraphics[width=0.85\columnwidth]{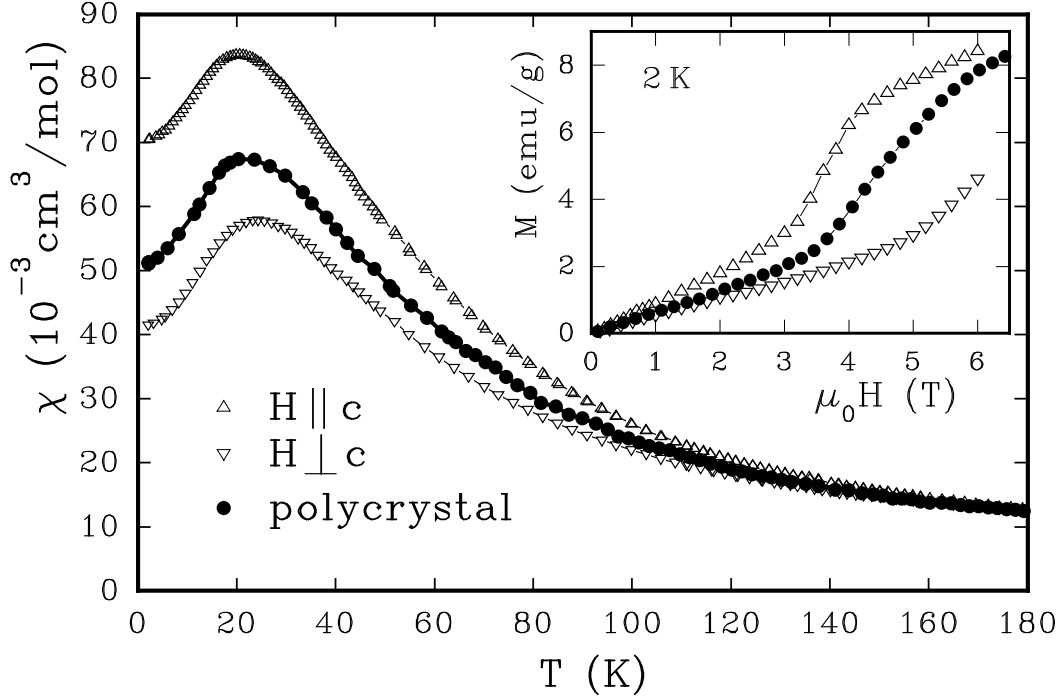}
\caption{Dc magnetic susceptibility of $\chi(T)$ of LaCo$_9$Si$_4$ 
measured at 1\,T on a large randomly oriented polycrystalline 
sample (filled symbols) and on a c-axis oriented textured specimen
with $H||c$ and $H\bot c$ (open symbols); inset: the corresponding
isothermal magnetization curves $M(H)$ measured at 2\,K.}
\label{fig1}
\end{figure}

The isothermal magnetization $M(H)$ measured on a bulk polycrystal 
LaCo$_9$Si$_4$ in a 15\,T VSM is shown in Fig.~2a 
and as Arrott plots ($M^2$ vs. $H/M$) in Fig.~2b.  
The 2 and 20\,K results indicate itinerant electron metamagnetism (IEMM),
i.e.\ a phase transition to a field induced FM state at $\mu_0H_c\sim 3$--6\,T 
with a typical S-shape  of the Arrott plot. The absence of any hysteretic 
behaviour (within the resolution of both, VSM and SQUID)
indicates that the  transition is on the verge to second order and  
may be connected with the vicinity to a field induced quantum critical point.  
The extrapolated "spontaneous" magnetization 
of the field induced FM state $M_0$ is about 0.9\,$\mu_B$/f.u.\ [see the dotted
line in Fig.~2b yielding $M_0^2\sim 0.8$\,$(\mu_B/$f.u.$)^2$]. 
\begin{figure}[]
\includegraphics[width=0.83\columnwidth]{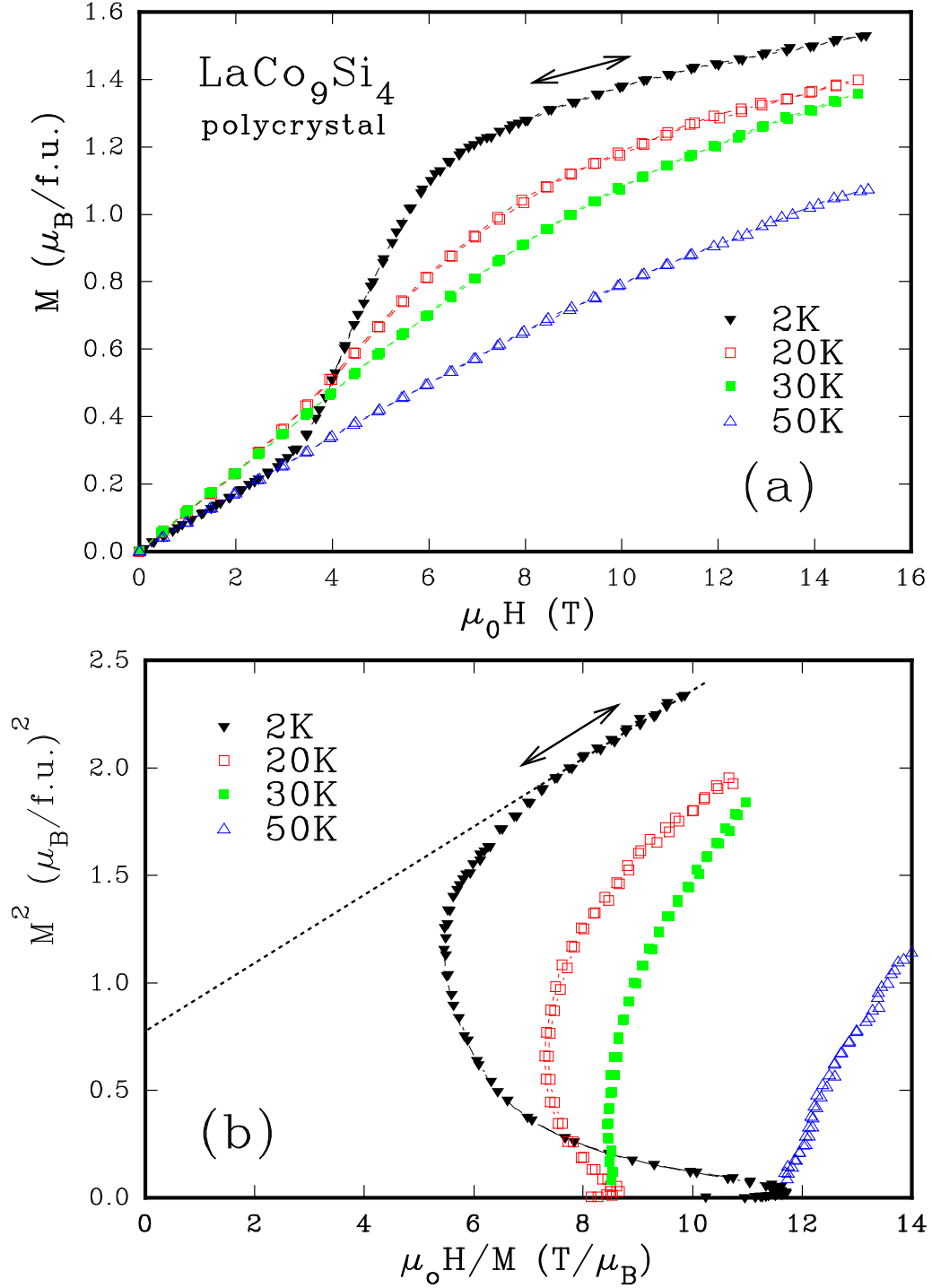}
\caption{(a) Isothermal magnetization $M(H)$ of polycrystalline LaCo$_9$Si$_4$; 
(b) Arrott plot: $M^2$ versus $H/M$.} 
\label{fig2}
\end{figure}
The considerable width of the metamagnetic critical field $\mu_0H_c\sim 3$--6\,T 
obtained on a polycrystalline sample is due to the random orientation of 
crystallites with respect to the external magnetic field. 
For oriented crystals the IEMM transition is observed at about 3--4\,T for 
$H||c$ and near 6\,T for $H\bot c$ (see inset of Fig.~1), i.e.\
the metamagnetic critical fields are inversely proportional to the 
corresponding susceptibilities $\chi_0 (H||c)$ and $\chi_0(H\bot c)$.
The almost infinite initial slope of the Arrott plot at 30\,K 
indicates the upper temperature limit of IEMM. 

\begin{figure}[]
\includegraphics[width=0.85\columnwidth]{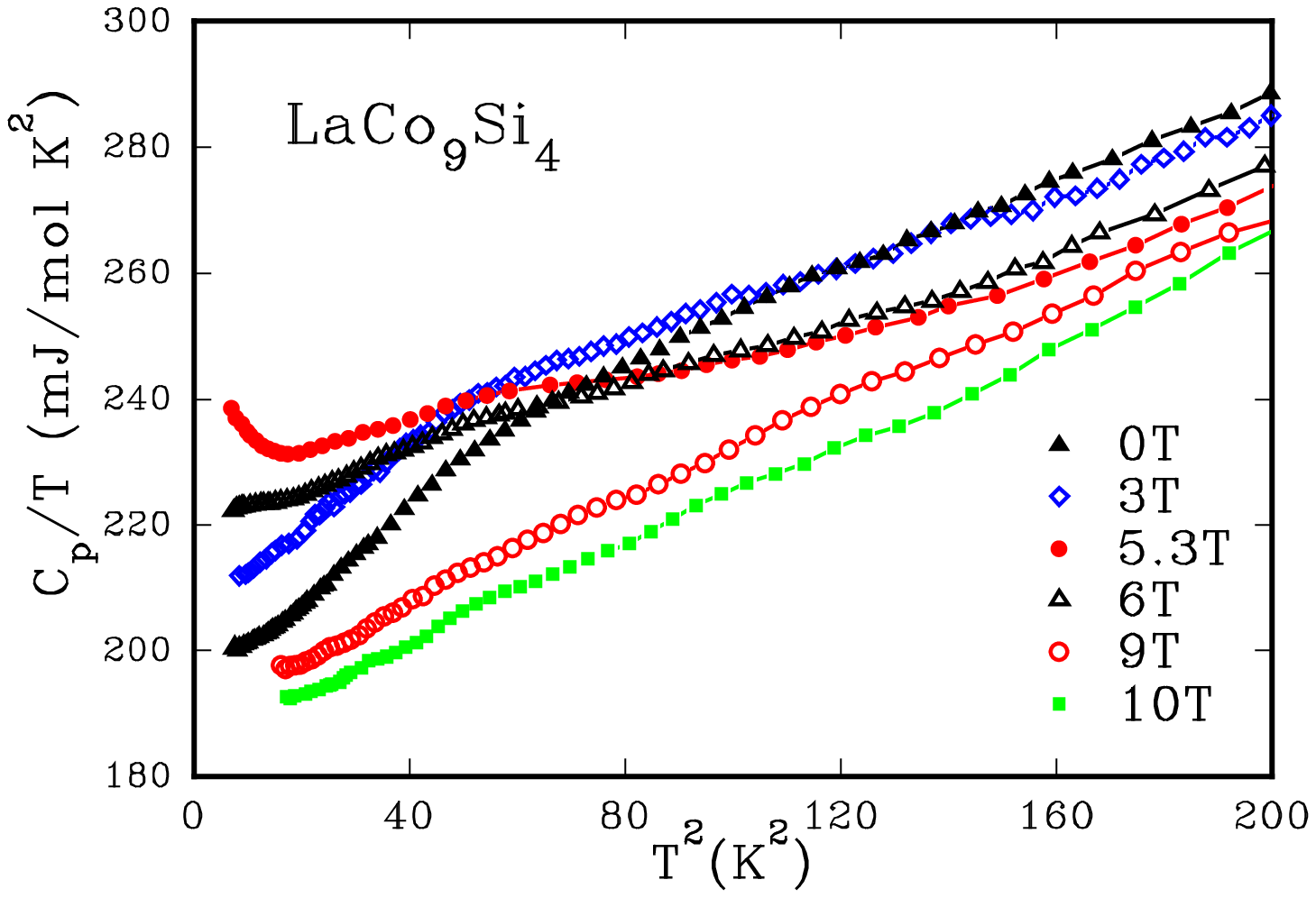}
\caption{Temperature dependent specific heat $C_p(T)$ of LaCo$_9$Si$_4$
displayed as $C/T$ vs. $T^2.$}
\label{fig3}
\end{figure}

The specific heat of polycrystalline LaCo$_9$Si$_4$ measured in external 
magnetic fields up to 10\,T (see Fig.~3) reveals an initial increase of the 
electronic specific heat coefficient $\gamma $ 
from 200\,mJ/mol\,K$^2$ at zero field to a maximum value of above 
230\,mJ/mol\,K$^2$ at 5.3\,T.  
Upon further increasing the magnetic field there is a 
significant reduction of the $\gamma $-value down to about 190 mJ/mol\,K$^2$ 
at 10\,T. 
The direct measurement of the magnetocaloric effect at about 3\,K 
(not shown for brevity) where $\Delta T(H)$ [approximately 
proportional to $-\Delta S(H)$] reveals a state of maximum entropy at 5.3\,T 
which defines a thermodynamic critical field $\mu_0H_c^T\simeq 5.3$\,T 
of the polycrystal being in close agreement with the weighted mean of 
$\frac{1}{3}\mu_0H_c||c+\frac{2}{3}\mu_0H_c\bot c\simeq 5.2$\,T.
The superposition of specific heat contributions of randomly oriented crystallites 
inhibits a straight forward analysis of the thermodynamic features of the critical state.
Nevertheless it is noteworthy, that below about 3\,K a low temperature upturn 
develops in $C/T$ just at $\mu_0H_c^T= 5.3$\,T being indicative for quantum 
criticality which may of course be limited to areas or spots on the Fermi 
surface. Similar observations indicating metamagnetic quantum criticallity
have been reported for CeRu$_2$Si$_2$ and Sr$_3$Ru$_2$O$_7$ (see
e.g.\ Refs.~\cite{flouquet,perry}) which is remarkable because of the
distinctly different mechanisms involved in each of these cases.

\begin{figure}[]
\includegraphics[width=0.82\columnwidth]{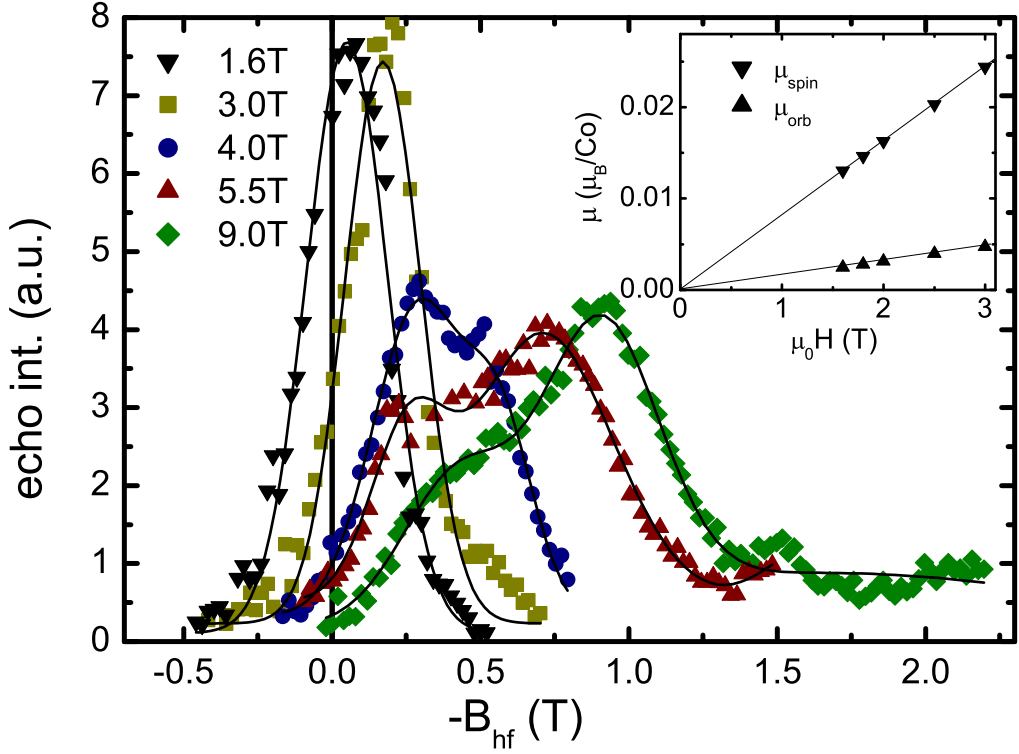}
\caption{NMR spin echo amplitude of LaCo$_9$Si$_4$ powder at 
7.5\,K vs.\ negative $B_{hf}$ for various applied fields  
as labeled; inset: spin and orbital moments below the critical
field, for higher field this evalution becomes impossible (see text).}
\end{figure}

The $^{59}$Co NMR spin-echo spectra of LaCo$_9$Si$_4$ at 7.5\,K have been 
measured in external fields between 1.5 and 9\,T. Neglecting dipolar and 
transferred field contributions the resonance frequency $f$ is 
determined by $f/\gamma =B_{hf}+\mu_0H$, where $\gamma=10.1$\,MHz/T is the 
gyromagnetic ratio of $^{59}$Co and $B_{hf}$ is the hyperfine field due to a magnetic 
moment $\mu$ in the electronic shell of the nucleus. 
\begin{figure}[]
\includegraphics[height=14.4cm]{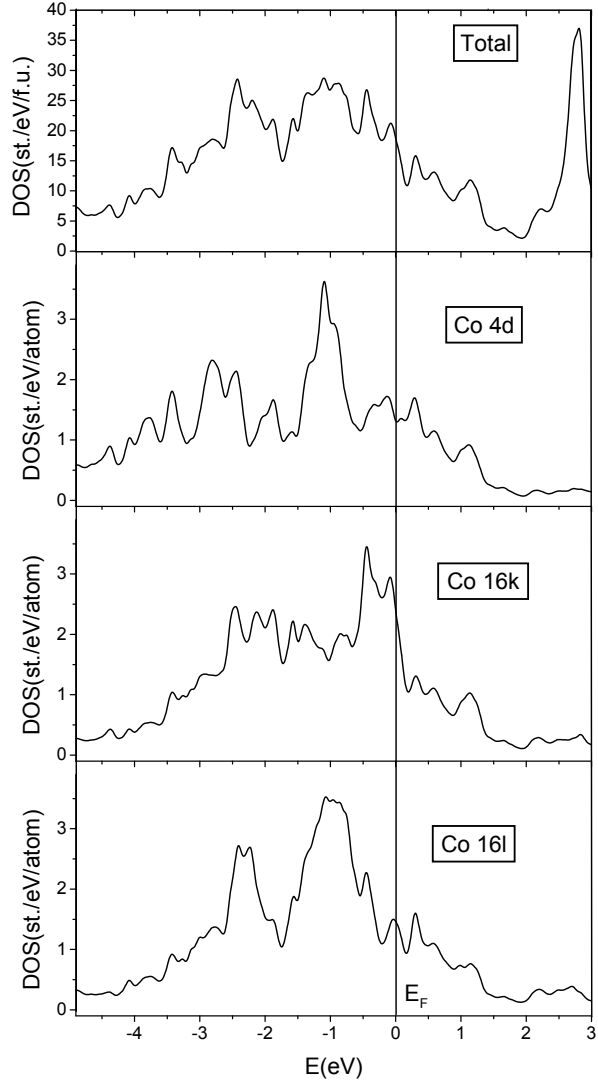}
\caption{Electronic density of states per unit cell (total) and
atom projected for the 3 different Co sites of LaCo$_9$Si$_4$.}
\label{fig4a}
\end{figure}
Figure~4 shows that even below the IEMM transition the considerable
inhomogeneous line width of $\Delta B_{hf}=0.25$\,T due to induced
moments does not allow to resolve the three crystallographic sites. At
$\mu_0H\geq 3.5$\,T a drastic line broadening indicates the IEMM and the
spectrum develops an unresolved splitting into two lines with a broad
shoulder to high internal fields. The inhomogeneous broadening of the
transition in the field range up to 6\,T is consistent
with the magnetic anisotropy discussed above. For $\mu_0H\geq 6$\,T the 
shape of the spectrum becomes again independent of the external field 
with a low intensity line at $-B_{hf}\simeq 0.22$\,T, a line of roughly 
three times this intensity at $-B_{hf}\simeq 0.75$\,T, and a broad shoulder 
in the range $-B_{hf}\sim $ 1--2.3\,T. 
From the line intensities and the band structure calculations discussed 
below we tentatively assign these structures to $4d$-, $16l$-, and $16k$-Co,
respectively.

Below the IEMM transition we estimate the mean Co spin and orbital 
moment $\mu_{spin}$ and $\mu_{orb}$  from $B_{hf}$ and the total magnetization 
(Fig.\,2) by solving $B_{hf}=\alpha\mu_{spin}+\beta\mu_{orb}$ and
$\mu =\mu_{spin}+\mu_{orb}$ with hyperfine coupling constants 
$\alpha =-12$\,T/$\mu_B$, $\beta =65$\,T/$\mu_B$~\cite{streever}. 
The result of this evaluation is shown in the inset 
of Fig.\,4 revealing a constant spin susceptibility of $9\times 10^{-3}$\,$\mu_B/$T\,Co 
and an orbital contribution of about $1.5\times 10^{-3}$\,$\mu_B/$T\,Co. 
This decomposition 
becomes impossible with the line broadening and unresolved splitting 
at higher field due to the unknown Co sublattice magnetizations.
Under the assumption of equal ratio $\mu_{spin}/\mu_{orb}$ for all
Co-sites we find with the line assignment above sublattice moments 
at 9\,T of 0.18, 0.09, and 0.04\,$\mu_B/$Co at $16k$-, $16l$-, and $4d$ sites,
respectively.

The ab-initio electronic structure calculations were performed employing the 
Full-Potential Linear-Augmented-Plane-Wave (FPLAPW) method~\cite{wimmer}. 
The effects of exchange and correlation were treated within the local density 
functional formalism including the general gradient approximation~\cite{wang}. 
The Brillouin zone integration was performed for 216 {\bf k}-points to achieve 
self consistency and for 1728 {\bf k}-points to determine the density of states 
(DOS) given in Fig.~\ref{fig4a} which shows the total DOS and the site projected 
DOS for the 3 different Co sites. 
From the total DOS it can be seen that the Fermi energy ($E_F$) lies in a strongly 
falling flank of a pronounced peak. 
Such a feature is known to be a necessary requirement for the occurrence of a 
metamagnetic phase transition~\cite{rhodes}. 
This peak in the total DOS stems from the contribution by the Co atoms at the 
$16k$ position, where the DOS at $E_F$ is so large to roughly fulfill the Stoner 
criterion. 
For the two other Co positions the DOS at $E_F$ is significantly lower. 
The Co atoms at the $16k$ positions form planes within the crystal structure which 
are separated from each other by almost non-magnetic {\it spacers} causing a 
narrowing of the $d$-band. 
To study the metamagnetic behavior Fixed Spin Moment (FSM) calculations were 
performed in the same fashion as it was done for YCo$_2$ earlier~\cite{schwarz}. 
At the experimental lattice constant we obtain a ferromagnetically ordered ground 
state. At slightly reduced volume a metamagnetic behaviour develops.
The magnetic moments obtained in the IEMM state are 0.3, 0.07, and 0.02\,$\mu_B/$Co 
at $16k$-, $16l$-, and $4d$ sites, respectively, in fair agreement with the 
experiments. 
The total energy versus magnetic moment curves show the typical behaviour like 
the archetypal itinerant metamagnet YCo$_2$~\cite{schwarz}.

The total DOS of LaCo$_9$Si$_4$ shown in Fig.~\ref{fig4a} yields at $E_F$ about 
19\,states/eV\,f.u.\ which implies a $T$-linear specific contribution 
$\gamma_b\sim 45$\,mJ/mol\,K$^2.$ The latter value is just about one fourth of
the experimental value, $\gamma\simeq $ 200 mJ/mol\,K$^2$, revealing a 
spin-fluctuation mass enhancement $\lambda_{spin}\sim 3.3$
which indicates strongly exchange enhanced Pauli paramagnetism.  
The mass enhancement of about three 
is also in line with the observed instability towards itinerant
ferromagnetism and compares 
well with related IEMM systems like YCo$_2$, however, with a 
significantly smaller critical field in the case of LaCo$_9$Si$_4$ which is 
the lowest value ever found for rare earth intermetallic compounds.

To summarize, LaCo$_{9}$Si$_{4}$ is a true ternary compound with a fully
ordered tetragonal LaFe$_{9}$Si$_{4}$-type structure which exhibits
itinerant electron metamagnetism with an anisotropic critical field 
$\mu_0H_{c}||c\sim 3.5$\,T and $\mu_0H_{c}\bot c\sim 6$\,T. The
anisotropy of the latter is in clear correspondence with the anisotropy of
the paramagnetic susceptibility and arises from small but finite 
Co-orbital moments resolved by NMR. 
Both the bandstructure calculations and NMR reveal significantly different
contributions from the three Co sites where Co at the $16k$-positions are
found to be responsible for the metamagnetic transition. This phase
transition appears to be on the verge to second order and may be connected with
a field induced quantum critical point. The vicinity to quantum criticality is
supported by the magnetocaloric effect revealing an entropy maximum just at the 
critical field of 5.3\,T (for bulk specimens) together with a small low
temperature upturn of $C/T$ at $H_C^T$, which resembles the metamagnetic
criticality of CeRu$_2$Si$_2$ and Sr$_3$Ru$_2$O$_7.$
Despite the differences in spin and electronic dimensionality common
trends emerge as described by Millis {\sl et al.}~\cite{millis} for
metamagnetic quantum criticallity: the finite temperature peak in the 
susceptibility, the maximum of entropy and $\gamma $ at the critical 
field and the paramagnetic ground state that should, according to
band structure calculations, be ferromagnetic.  
Accordingly, LaCo$_9$Si$_4$ and related novel intermetallics with 1:9:4 
stoichiometry are fascinating systems to study the interrelations of spin 
fluctuation behaviour, IEMM, and quantum criticality in structures with 
full translational symmetry. 
They might be promising candidates to search for the appearance and/or 
coexistence of superconductivity and itinerant electron ferromagnetism.

The work was supported by the Austrian Science Foundation under
project P-15066-Phy and P-15700-Phy. The work by M. El-H. was funded by the
\"OAD.  S.K. acknowledges support form the Austrian Ministry of Science 
Contract Number: GZ 650.758/1-VI/2/2003.

\end{document}